\documentclass[aps,pre,twocolumn,groupedaddress,showkeys,showpacs]{revtex4-1}

\usepackage{graphicx} 
\usepackage{amsmath}
\usepackage{subfigure}
\usepackage{enumerate}
\usepackage{hyperref}

\newcommand{\dd}[1]{\mathrm{d}{#1}}
\newcommand{\KL}[2]{D_\mathrm{KL}(#1\|#2)}
\renewcommand{\KL}[2]{D(#1\|#2)}
\newcommand{\hide}[1]{{}}

\begin{document}


\title{Ensemble annealing of complex physical systems}

\author{Michael Habeck}
\email[]{mhabeck@gwdg.de}
\affiliation{Max Planck Institute for Biophysical Chemistry, Am Fa{\ss}berg 11
}
\affiliation{Felix Bernstein Institute for Mathematical Statistics in the Biosciences, University of G\"ottingen, Goldschmidtstrasse 7, 37077 G\"ottingen, Germany}

\date{\today}

\begin{abstract}
Algorithms for simulating complex physical systems or solving difficult optimization problems often resort to an {annealing} process. Rather than simulating the system at the temperature of interest, an annealing algorithm starts at a temperature that is high enough to ensure ergodicity and gradually decreases it until the destination temperature is reached. This idea is used in popular algorithms such as parallel tempering and simulated annealing. A general problem with annealing methods is that they require a temperature schedule. Choosing well-balanced temperature schedules can be tedious and time-consuming. Imbalanced schedules can have a negative impact on the convergence, runtime and success of annealing algorithms. This article outlines a unifying framework, {ensemble annealing}, that combines ideas from simulated annealing, histogram reweighting and nested sampling with concepts in thermodynamic control. Ensemble annealing simultaneously simulates a physical system and estimates its density of states. The temperatures are lowered not according to a prefixed schedule but adaptively so as to maintain a constant relative entropy between successive ensembles. After each step on the temperature ladder an estimate of the density of states is updated and a new temperature is chosen. Ensemble annealing is highly practical and broadly applicable. This is illustrated for various systems including Ising, Potts, and protein models. 
\end{abstract}

\pacs{}
\keywords{free energy; density of states; annealing; histogram reweighting;  Monte Carlo simulation; replica-exchange Monte Carlo}

\maketitle
\section{Introduction}
Simulational science often involves the generation of configurations from high-dimensional probability distributions as well as the computation of ensemble averages and normalization constants. Numerous applications in statistical physics, biomolecular simulation and Bayesian inference illustrate the ubiquitous need for efficient sampling methods. Challenges are posed by the complexity of the system, its shear size, slow convergence and non-ergodicity. 

To address these challenges, algorithms that work with modified versions of the system have been proposed. One idea is to simulate the system at multiple temperatures and utilize the enhanced flexibility at higher temperatures to avoid local free-energy minima at lower temperatures. This idea is the basis of sampling algorithms such as replica-exchange Monte Carlo \cite{Swendsen86} and simulated tempering \cite{Marinari92} but also used in popular optimization algorithms such as simulated annealing \cite{Kirkpatrick83}. 

Parallel tempering (PT) \cite{Geyer91}, for example, considers a family of canonical ensembles at different temperatures. The ensembles are simulated independently and occasional swaps of configurations between ensembles at nearby temperatures allow the simulation to escape from metastable states. From a PT simulation thermodynamic quantities such as free energies and heat capacities can then be computed with high accuracy. But the success and convergence of a PT run depends critically on the choice of the temperature schedule. To choose a good temperature schedule can be highly non-trivial, especially for systems undergoing phase transitions. A well-balanced schedule entails overlap between ensembles at neighboring temperatures. This means that we have to use more and more replicas with increasing system size because energy is extensive \cite{Predescu05}. Moreover, PT explores temperature space on a fixed ladder. If we want to use multiple temperatures or control parameters as in multi-dimensional PT \cite{Sugita00} we are suffering from the curse of dimensionality. Another source of inefficiency is the fact that configurations at high temperatures are constantly being produced but no longer needed once the simulation has converged. 

Multi-canonical sampling algorithms \cite{Berg92} are a powerful alternative to annealing methods. Rather than utilizing a temperature parameter to modify the system, multi-canonical algorithms draw configurations from an ensemble whose weight is inversely proportional to the density of states (DOS) of the system, such that ideally the energy histogram will be constant. However, this requires that we know the DOS {before} the actual simulation, which is rarely the case. 

The Wang-Landau (WL) algorithm \cite{Wang01} is an ingenious variant of multi-canonical sampling that sidesteps this problem. The unknown density of states is estimated in the course of a WL run, configurational samples are generated as a by-product. The fact that the correct DOS should produce a flat energy histogram can be used to monitor the convergence of the method. By gradually decreasing the learning rate, the simulation is stabilized and converges. 

WL sampling has originally been developed for discrete systems \cite{Wang01}. Its direct extension to large or continuous systems requires choosing an energy range and binning. But there might be forbidden energy levels that cannot be visited, in which case the corresponding bins remain empty and the energy histogram will never be flat. These problems are aggravated for multi-dimensional DOS over more than one macrovariable because the number of bins grows exponentially in the number of macrovariables. In that case flatness of the energy histogram ceases to be a useful convergence criterion and must be replaced by other criteria \cite{Zhou06}. To apply these modifications in practice remains a challenge and involves parameter tweaking. 

This article proposes an algorithm, {\em ensemble annealing}, that solves these issues and produces samples and an estimate of the DOS. Ensemble annealing is inspired by the nested sampling method for Bayesian computation \cite{Skilling06} and can be viewed as a generalization of nested sampling to the canonical or other ensembles. The algorithm applies both to discrete and continuous systems. In contrast to simulated annealing or parallel tempering, ensemble annealing constructs an optimal temperature protocol adaptively and has only few algorithmic parameters. 
\section{Ensemble annealing}
Ensemble annealing is a sequential algorithm that steps through iterations denoted by $k$. $N$ non-interacting particles or walkers are employed to explore a series of ensembles typically starting in a high temperature ensemble, then cycling through ensembles at lower and lower temperatures, until the destination ensemble is reached. For each ensemble, the walkers produce configurations $x_{kn}$ where $n=1,\ldots,N$. In contrast to other annealing and tempering methods only the start and final ensemble have to be chosen. The intermediate ensembles are found during the simulation by placing them such that a constant overlap between successive ensembles is maintained. To implement this approach, we need to agree on various concepts, mainly what kind of ensembles will be considered, and how to measure distances between ensembles. 
\subsection{Ensembles}
Let us denote the target ensemble from which we aim to generate configurations $x$ by $p(x)$. Often $p(x) = \pi(x)\, e^{-E(x)}$ with energy $E(x) = - \ln p(x)$. In Bayesian inference, for example, $\pi(x)$ denotes the prior distribution and $E(x)$ corresponds to negative log likelihood. In a physical simulation of particles in a box, $\pi(x)$ will be uniform over the box and $E$ will be the interaction energy between all particles. Note that in practice both $\pi(x)$ and $p(x)$ are often unnormalized. 

To draw configurations from $p(x)$ we consider a series of ensembles
\begin{equation}\label{eqn:family}
p_k(x) = \frac{1}{c_k}\, q_k[E(x)]\, \pi(x)
\end{equation}
where $c_k = \int q_k[E(x)]\, \pi(x)\, \dd{x}$ normalizes the $k$-th ensemble. Here, we assume that ensemble $p_k$ depends on the configuration only through the macrovariable $E(x)$, but the method works also for more general ensembles. Typically $q_k(E) = q(E;\lambda_k)$ where $q(E;\lambda)$ is a parameterized family and $\lambda$ a protocol parameter. The distributions $q_k(E)\ge0$ are intermediate helper or bridging distributions. In case of the canonical ensemble, configurations are weighted by the Boltzmann factor 
\begin{equation}\label{eqn:canonical}
q_k(E) = \exp\{-\beta_k E\}
\end{equation}
where $\beta_k$ is the inverse temperature and $c_k$ is the partition function. 

Obviously the canonical ensemble is a widespread choice in annealing methods, but it might also be worthwhile to consider other ensembles. For example, in parallel tempering the use of the Tsallis ensemble 
\begin{equation}\label{eqn:tsallis}
q_k(E) = \frac{1}{\{1 +  \beta (\alpha_k-1) (E - E_{\min})\}^{1/(\alpha_k-1)}}
\end{equation}
with control parameter $\alpha_k \ge 1$ \footnote{Normally, the control parameter of the Tsallis distribution is called $q_k$. Because we already use $q_k$ for the bridging ensemble, we denote the Tsallis parameter by $\alpha_k$.}, inverse temperature $\beta$ and minimum energy $E_{\min} \le E(x)$ has been proposed \cite{Hansmann97}. A multi-parameter combination of the Boltzmann and Tsallis ensemble is used in complex Bayesian data analyses \cite{Habeck05} to independently control the prior density and the likelihood function. 

Another ensemble that is of potential interest is the Fermi distribution
\begin{equation}\label{eqn:fermi}
q_k(E) = \frac{1}{1 + \exp\{\beta_k(E - \epsilon_k)\}}
\end{equation}
which has two control parameters: the inverse temperature $\beta_k$ and an energy cutoff  $\epsilon_k$. The zero-temperature Fermi ensemble approaches a stepfunction, i.e. configurations with energies greater than $\epsilon_k$ are assigned zero probability:
\begin{equation}\label{eqn:nested}
q_k(E) = \Theta(\epsilon_k - E)
\end{equation}
where $\Theta(\cdot)$ is the Heaviside step function. This ensemble is used in the nested sampling method for Bayesian computation \cite{Skilling06} and also related to the microcanonical ensemble \cite{Ray91,Martin-Mayor07}:
\begin{equation}\label{eqn:ray}
q_k(E) = \Theta(\epsilon_k - E)\, (\epsilon_k - E)^{d/2-1}
\end{equation} 
where $d$ is the dimension of configuration space (number of configurational degrees of freedom) and $\epsilon_k$ the total energy of the system (potential plus kinetic energy). 

Note that the target ensemble $p(x)$ does not necessarily need to be a member of the bridging family, i.e. there might be no $p_k(x)$ such that $p(x) \propto p_k(x)$, which is the case, for example, in nested sampling and the microcanonical ensemble. 

The density of states (DOS) over the prior or reference distribution $\pi(x)$ is defined as
\begin{equation}\label{eqn:dos}
g(E) = \int \delta(E - E(x))\, \pi(x)\, \dd{x}
\end{equation}
with $\delta(\cdot)$ denoting the delta function. With the help of the DOS it is straightforward to compute how the energies are distributed in the intermediate ensembles:
\begin{equation}\label{eqn:ensemble_energy}
p_k(E) = \int \delta(E - E(x))\, p_k(x)\, \dd{x} = \frac{1}{c_k}\, g(E)\, q_k(E)
\end{equation}
where $p_k(E)$ is a one-dimensional distribution. 
\subsection{Relative entropy}
When choosing the intermediate distributions that bridge between the initial and final ensemble, it is essential to control the ``distance'' or overlap between successive ensembles $p_k(x)$ and $p_{k+1}(x)$. We use the Kullback-Leibler (KL) divergence  \cite{Kullback51} or relative entropy
\begin{equation}\label{eqn:relativeentropy}
\KL{p}{q} = \int p(x) \ln[ p(x) / q(x)]\, \dd{x}
\end{equation}
for this purpose. The relative entropy satisfies the Gibbs inequality $\KL{p}{q} \ge 0$ with equality only if $p$ and $q$ are identical. Therefore the Kullback-Leibler divergence qualifies as an ``entropic distance'' between ensembles $p$ and $q$. However in contrast to a true distance the KL divergence is not symmetric under interexchange of $p$ and $q$. It is only well-defined if $q$ is ``broader'' than $p$, i.e. if the support of $p$ is contained in the support of $q$, and therefore a \textit{directed divergence}. In information theory, the KL divergence is used to quantify information \textit{gain}. 

Let us now consider the relative entropy between two members $p_k$ and $p_l$ of the family of bridging distributions [Eq. (\ref{eqn:family})]. With the help of the DOS we can reduce the high-dimensional configurational integral [Eq. (\ref{eqn:relativeentropy})] to a one-dimensional integral over the energies:
\begin{eqnarray}\label{eqn:KL}
\nonumber \KL{p_{k}}{p_{l}} &=&\int p_k(E) \ln\left\{\frac{q_k(E)\, c_{l}}{q_{l}(E)\, c_{k}} \right\}\dd{E} \\
&=& \langle \ln( q_k / q_l )\rangle_k - \ln(c_k/c_l)
\end{eqnarray}
where $\langle \cdot \rangle_k$ denotes an average over the $k$-th ensemble $p_k$. For the canonical ensemble [Eq. (\ref{eqn:canonical})] the relative entropy reduces to
\begin{equation}\label{eqn:KLcanonical}
\KL{p_{k}}{p_{l}} = (\beta_l - \beta_k) \langle E \rangle_{\beta_k} + \beta_k F(\beta_k) - \beta_l F(\beta_l)
\end{equation}
where $F(\beta) = - \beta^{-1} \log c(\beta)$ is the free energy at inverse temperature $\beta$. 

Throughout this article, we will use the relative entropy to measure the distance between ensembles $p_{k}$ and $p_{l}$. Other measures that quantify the overlap between different ensembles might also be useful. For example, we could use the exchange rate of a parallel tempering simulation
\begin{eqnarray}
\nonumber R(\beta_k,\beta_l) &=& \frac{1}{c_k c_l}\int \min\left\{q_k(E_1)q_l(E_2),q_k(E_2)q_l(E_1)\right\} \\ &&\qquad\qquad\times\,\,  g(E_1) g(E_2) \, \dd{E_1}\dd{E_2}
\end{eqnarray}
as a measure to compare ensembles. The Jensen-Shannon divergence \cite{Lin91}, a symmetrized version of the relative entropy, has been used in thermodynamic control \cite{Crooks07b}. The Hellinger distance \cite{Gelman98} is a widespread distance used mainly in statistics and may also provide a useful measure for comparing ensembles. In this article, however, we have not explored measures for comparing ensembles other than the relative entropy. 

Given a continuous bridging family, the optimal annealing protocol would involve infinitely many steps (adiabatic annealing). We want to reach the target ensemble in finitely many steps but produce intermediate ensembles that have a fixed and finite relative entropy $D$. We will later see that for small $D$ this amounts to cooling with constant thermodynamic speed. As we move from ensemble $p_k$ to the next ensemble $p_{k+1}$ we need to evaluate their relative entropy $\KL{p_{k+1}}{p_k}$. Equation (\ref{eqn:KL}) shows that this involves the computation of ensemble averages as well as the estimation of free energy differences. These are challenging computational problems, which can be solved by the methods outlined in the next subsection.  
\subsection{Estimation of the relative entropy}
Because the relative entropy [Eq. (\ref{eqn:KL})] both involves the normalization constants $c_k$, $c_l$ as well as an ensemble average, it is computationally challenging to evaluate $\KL{p_k}{p_l}$ accurately. However, if we know the density of states $g(E)$, the configurational integrals can be reduced to low-dimensional integrals. Therefore, ensemble annealing estimates $g(E)$ during the course of the simulation, similar to the Wang-Landau method \cite{Wang01} or nested sampling \cite{Skilling06}. The estimation of the DOS relies on histogram methods \cite{Ferrenberg89,Habeck12b}. 

If we work with $N$ non-interacting walkers at the $k$-th iteration, the configurations are denoted by $x_{kn}$ (i.e. the first index indicates the ensemble, whereas the second index enumerates the walkers). At each ensemble annealing iteration $k$, a non-parametric estimate of the DOS 
\begin{equation}\label{eqn:dosestimate}
g^{(k)}(E) = \sum_{n=1}^N \sum_{k'=0}^{k-1} g^{(k)}_{k'n} \, \delta(E-E_{k'n})
\end{equation}
is updated where $E_{kn} = E(x_{kn})$ are the energies of the visited configurations. The discrete DOS $g_{ln}^{(k)}$ assigns a weight to every configuration $x_{ln}$ that has been generated by the walkers during the entire simulation up to the current ensemble $p_k$. That is, the vector of all weights expands in each iteration and is constantly updated (which is indicated by the superscript). 

With the help of the estimated DOS it is straightforward to compute the relative entropy between two ensembles $p_k$ and $p_l$:
\begin{equation}\label{eqn:KLestimate}
\KL{p_k}{p_l} \approx \sum_{k'=0}^{k-1} \sum_{n=1}^N \frac{g_{k'n}^{(k)}\, q_k(E_{k'n})}{c_k} \ln \frac{q_k(E_{k'n})}{q_l(E_{k'n})} - \ln( c_k / c_l)
\end{equation}
where
\begin{equation}\label{eqn:Zestimate}
c_k \approx \sum_{k'=0}^{k-1} \sum_{n=1}^N g_{k'n}^{(k)}\, q_k(E_{k'n}).
\end{equation}
These relations are used in histogram methods for estimating free energy differences \cite{Habeck12,Habeck12b}. The weights are obtained using the histogram iterations
\begin{equation}\label{eqn:wham}
g_{k'n}^{(k)} \propto \left(\,\, \sum_{l=0}^{k-1}  q_{l}(E_{k'n}) / c_{l} \right)^{-1} 
\end{equation}
in which each update of the weights $g_{k'n}^{(k)}$ is followed by their normalization and a re-evaluation of the partition functions $c_{k'}$ according to Eq. (\ref{eqn:Zestimate}). We start the iteration from the previous DOS estimate (setting the weights of the new states $x_{kn}$ to zero), which speeds up the convergence of the histogram iterations. 

\subsection{Initialization and equilibration of ensembles}
The estimated DOS serves two purposes: First, to estimate the relative entropy between two ensembles reliably; second, to initialize the walkers to sample the next ensemble by recycling configurations that have been generated previously, which are then equilibrated in the new ensemble. In the $k$-th ensemble annealing iteration, ensemble $p_k(x)$ is approximated by
\begin{equation}\label{eqn:approxensemble}
p_k(x) \approx \frac{1}{c_k} \sum_{k'=0}^{k-1} \sum_{n=1}^N g_{k'n}^{(k)} q_k(E_{k'n}) \delta(x - x_{k'n}).
\end{equation}
We use this approximation to generate $N$ initial states for the walkers by the following scheme: First, we draw an energy level according to the probability $g_{k'n}^{(k)} q_k(E_{k'n}) / c_k$. Second, we randomly pick one among all configurations that map to the energy drawn in the first step. In continuous systems, it is very unlikely that two configurations were generated that have exactly matching energies. However, in discrete systems such as the two-dimensional Ising model there are only finitely many energy levels. In this case, we can speed up the DOS estimation [Eqs. (\ref{eqn:Zestimate}) and (\ref{eqn:wham})] by working with histograms as explained in \cite{Habeck12b}. Due to the limitation of the approximation (\ref{eqn:approxensemble}), the $N$ recycled states need to be equilibrated in the correct ensemble [Eq. (\ref{eqn:family})] $p_k(x) \propto \pi(x) q_k(E(x))$ using Monte Carlo or molecular dynamics simulations.
\begin{figure*}
{
\centerline{\resizebox{0.995\textwidth}{!}{\includegraphics{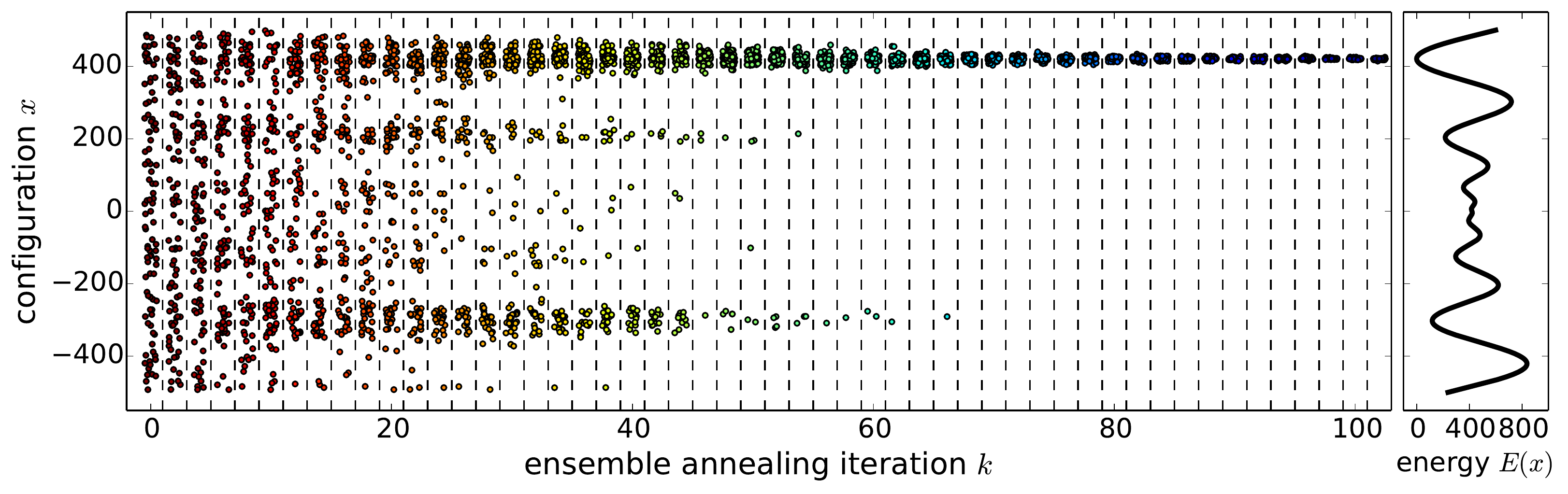}}}
}
\caption{(Color online) Ensemble annealing of a one-dimensional particle in a Schwefel potential $E(x) = -x\, \sin(\sqrt{|x|})$ (shown on the right) using 100 walkers. Every stripe marked by a dashed boundary shows the configurations of the walkers after the equilibration step. A random number has been added to the x-coordinates (iteration index) for better visualization. All Particles within each stripe experience the same temperature during equilibration. }\label{fig:schwefel}
\end{figure*}
\subsection{Algorithm}
We have now all tools at hand to formulate the ensemble annealing algorithm. Ensemble annealing is an adaptive sequential Monte Carlo algorithm. The main parameters are the number of walkers $N$ and the relative entropy $D$ between successive ensembles $p_k$ and $p_{k+1}$. Choosing ensembles with a constant relative entropy ensures that the annealing process proceeds at a constant thermodynamic speed. Iteration $k$ comprises the following steps:
\begin{enumerate}[(i)]
\item Initialization: Using the current estimate of the DOS $g^{(k)}(E)$ [Eq. (\ref{eqn:dosestimate})], the particles are initialized by drawing $N$ energies $E_{kn}'$ from $q_k(E)$ [Eqs. (\ref{eqn:ensemble_energy}) and (\ref{eqn:dosestimate})] and finding the corresponding configurations $x_{kn}'$ by a simple lookup in the energy table such that $E(x_{kn}') = E_{kn}'$. Because $g^{(k)}$ is only an approximation, the initial states will not be equilibrated.
\item Equilibration: The states are equilibrated in the new ensemble $p_{k}(x)$ by running Monte Carlo or molecular dynamics simulations starting from $x_{kn}'$ and producing new states $x_{kn}$. The new configurations and energies $E_{kn}$ are added to the pool of all states visited so far.  
\item DOS estimation: A new estimate of the DOS, $g^{(k+1)}$, is computed from all energies and temperatures using non-parametric histogram reweighting \cite{Habeck12b}. To speed up the convergence, the previous DOS estimate is used to initialize the iterations. 
\item Annealing: The next ensemble $p_{k+1}$ is adjusted such that it has a desired relative entropy $D$ with respect to the current ensemble $p_k$, i.e. $p_{k+1}$ satisfies $\KL{p_{k+1}}{p_{k}} = D$.
\end{enumerate}
The algorithm has only few parameters, namely the initial and final ensemble, the number of walkers $N$ and the target relative entropy $D$ between successive ensembles. Evidently, $D$ determines the cooling or compression rate. For smaller $D$ the overlap between successive ensembles is larger and the annealing progresses more slowly. If we allow $D$ to be large, we anneal faster but risk to fail to equilibrate. 
\subsection{Application to the harmonic oscillator}\label{sec:harmoniccanonical}
Let us illustrate ensemble annealing for a simple system, the one-dimensional harmonic oscillator with  energy $E(x) = (x-x_0)^2 / 2$ and ground state $x_0$ in the canonical ensemble:
\[
p_k(x) \equiv p(x|\beta_k) = \sqrt{\frac{\beta_k}{2\pi}} \exp\left\{-\frac{\beta_k}{2} (x - x_0)^2 \right\}.
\]
The distance between two ensembles at inverse temperatures $\beta$ and $\beta'$, $\beta'\ge\beta$, is: 
\begin{equation}\label{eqn:oscillator}
\KL{\beta'}{\beta} = \frac{1}{2}\, [ (\beta/\beta') - 1 - \ln (\beta/\beta')]
\end{equation}
In this case the KL divergence depends only on the ratio between two successive temperatures. The constant relative entropy criterion yields a constant cooling rate $\rho = \beta/\beta', 0\le \rho \le 1$ which is determined by 
\begin{equation}\label{eqn:canonicalrate}
 \rho - \ln \rho = 2D + 1.
\end{equation}
This results in the geometric schedule $\beta_k = \beta_0\, \rho^{-k}$. For $D\rightarrow 0$  we reach the adiabatic limit of infinitely slow cooling since $\rho\rightarrow 1$. Geometric schedules have been proposed for optimal simulated annealing \cite{Kirkpatrick83}. 

Alternatively, we could consider the ground state a control parameter, $\beta = x_0$, and let $\pi(x) \propto \exp\{-x^2/2\}$, $E(x) = -x$ with $\ln Z(\beta) = \beta^2/2$ and $\langle E \rangle = - \beta$. The relative entropy is now according to Eq. (\ref{eqn:KLcanonical}):
\[
\KL{\beta'}{\beta} = -\frac{(\beta')^2}{2} + \frac{\beta^2}{2} - (\beta-\beta') \beta' = \frac{1}{2} (\beta' - \beta)^2.
\] 
Constant steps in the relative entropy lead to a schedule that is linear in the inverse temperature: $\beta_k = \beta_0 \pm \sqrt{2 D}\, k$. That is, the ground state is shifted either in the positive or the negative direction depending on the targeted ground state. 

These examples highlight that it is not sufficient to prescribe the relative entropy to choose the next ensemble. We must also impose some sense of directionality in order to shift the ensemble closer to the target ensemble. This will become particularly important in multi-dimensional annealing. 
\section{Canonical ensemble}
We will now apply annealing of the canonical ensemble to various systems including discrete systems such as Ising and Potts models and a continuous protein model.
\subsection{One-dimensional example}
To illustrate ensemble annealing, we first apply it to a system with a one-dimensional configuration space and a rugged energy function $E(x) = - x \sin(\sqrt{|x|})$, the one-dimensional Schwefel function. We deliberately choose a large number of walkers for illustrative purposes ($N=100$); a smaller number of walkers would be sufficient in this one-dimensional example. At every iteration, equilibration is achieved by using a random walk Metropolis Monte Carlo scheme \cite{Metropolis57} consisting of 10 random steps drawn from a uniform distribution. The relative entropy is set to $D=10^{-3}$. Figure \ref{fig:schwefel} shows the configurations at the various temperatures obtained by the constant relative entropy criterion. We start at $\beta_\mathrm{initial} = 0$ and target a final inverse temperature $\beta_\mathrm{final} = 1$. As ensemble annealing progresses the walkers become more and more localized in the dominant modes of the target ensemble. The relative proportions are reproduced accurately. 

This example also suggests that it should be possible to prune the number of the walkers during the annealing process. In the course of annealing, the ensemble becomes more and more concentrated (as monitored by a decrease in the entropy $S_k = - \int p_k \ln p_k$), and fewer walkers are needed to explore and represent it. Using the Boltzmann relation $S = k \log W$ where $W$ is the number of accessible microstates, we could decrease the number of walkers in each iteration and thereby save computational resources. However, we have not explored this strategy further in this article. 
\subsection{Ising and Potts model}
\begin{figure}
\subfigure[\hspace*{0.2cm}Ensemble annealing of the $32\times{}32$ Ising model]{
\centerline{\resizebox{\columnwidth}{!}{\includegraphics{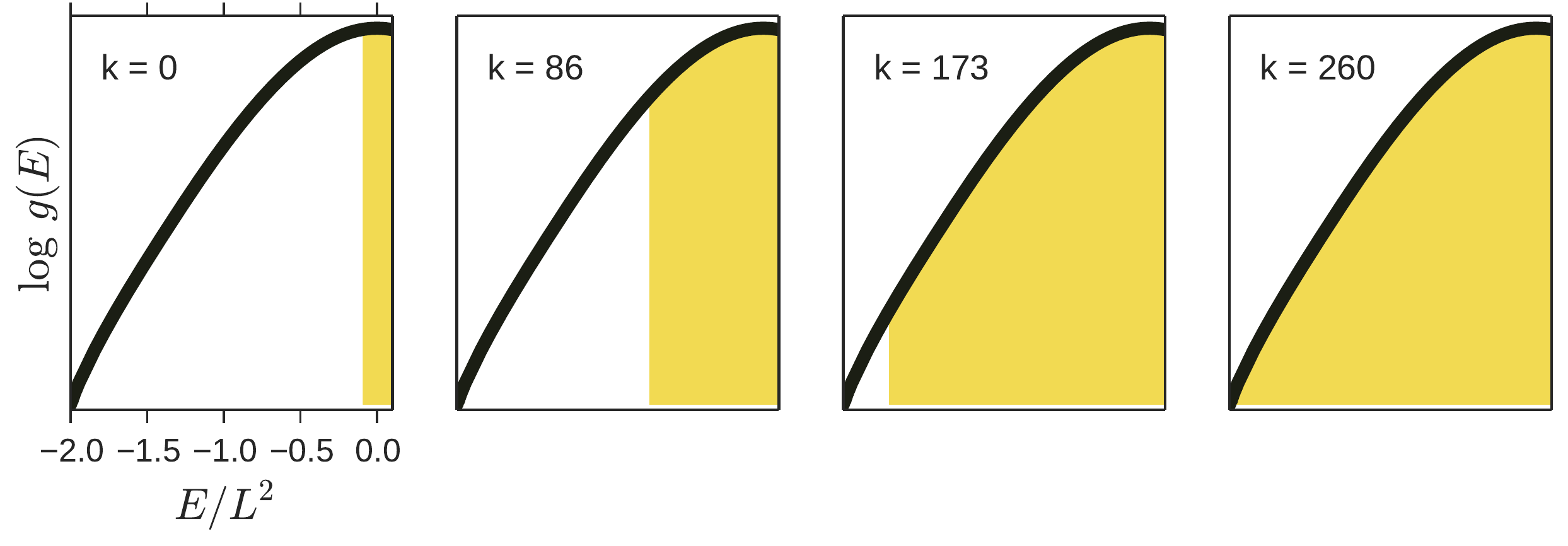}}}
}
\subfigure[\hspace*{0.2cm}DOS of the $32\times{}32$ Ising and Potts model]{
\centerline{\resizebox{\columnwidth}{!}{\includegraphics{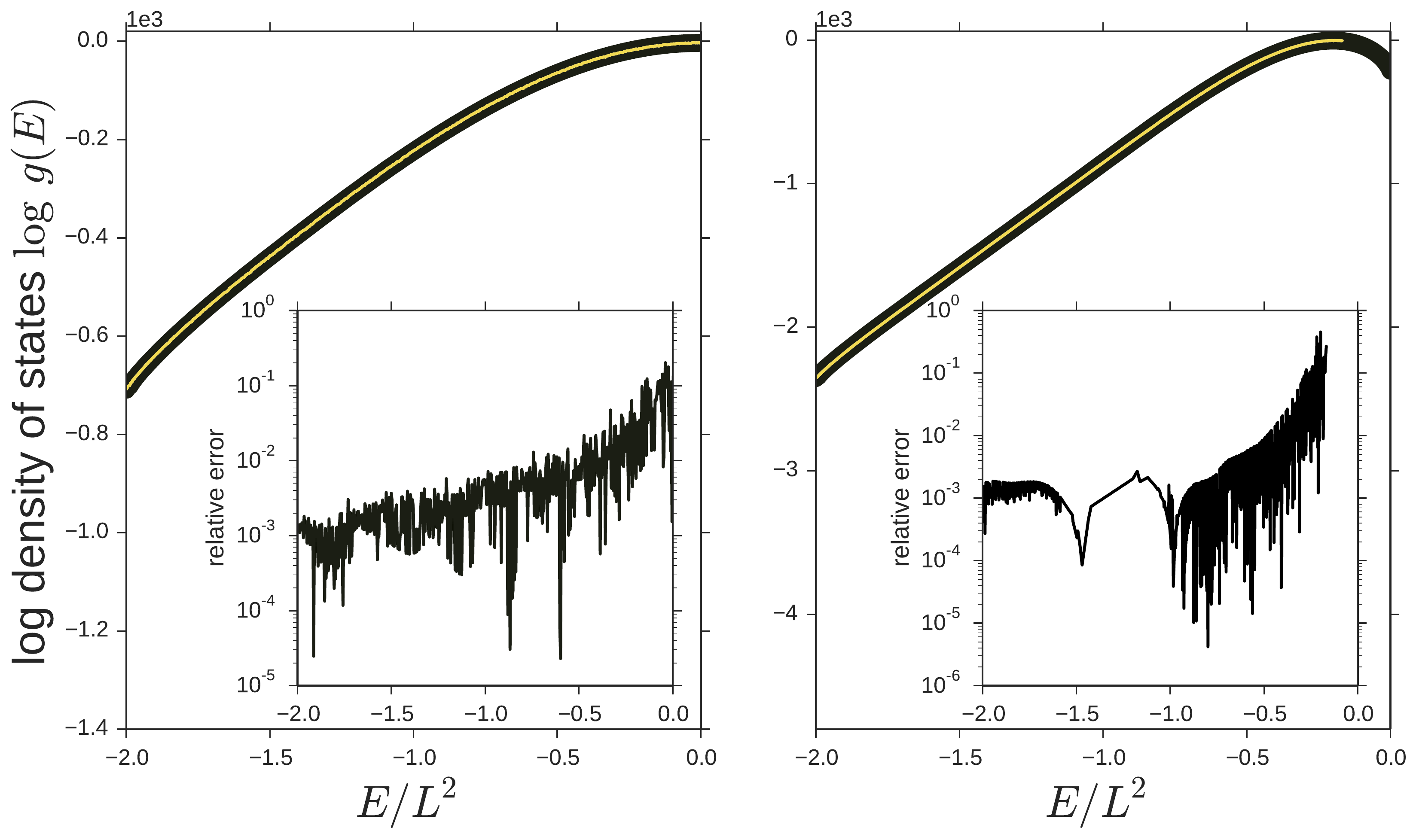}}}
}
\caption{(Color online) (a) Improvement of the estimated DOS during ensemble annealing. Four equally spaced stages of ensemble annealing were picked; the iteration index $k$ is indicated in the legend. The true DOS \cite{Beale96} is shown as a thick black curve; the estimated DOS as yellow [gray] area. (b) Estimated DOS for the Ising (left) and the ten state Potts model (right). Again the black line shows the true DOS and the yellow [gray] line the DOS produced during ensemble annealing. The insets show the relative error in the microcanonical entropy $s(E) = \ln g(E)$.}\label{fig:1}
\end{figure}

We now apply ensemble annealing to the two-dimensional Ising and Potts model. Figure \ref{fig:1} shows simulation results for the $32\times 32$ lattice. $N=10$ particles were used and the relative entropy was set to $D=10^{-2}$. The equilibration step consisted of Metropolis Monte Carlo runs that randomly select lattice sites and try to flip the spin (Ising model) or draw a random color (Potts model). Figure \ref{fig:1}(a) shows how the algorithm improves the initial DOS estimate. The algorithm starts with $N$ random spin configurations from which the initial DOS covering only a limited energy range is derived. As the algorithm proceeds, lower energy states are generated and the DOS  expands into the lower energy region. This process continues until the full energy range has been explored and a highly accurate estimate of the DOS is produced. The accuracy of the estimated DOS is illustrated in Fig. \ref{fig:1}(b). The estimation error can be as small as with WL sampling \cite{Wang01}. A similar accuracy is also obtained for the ten state Potts model which undergoes a first order phase transition. The DOS tends to be more accurate for the low energy states. In most situations this is desirable because one is primarily interested in the thermodynamic properties of the system at finite temperatures, at which the low energy states contribute most strongly. 
\begin{figure}
\centerline{\resizebox{\columnwidth}{!}{\includegraphics{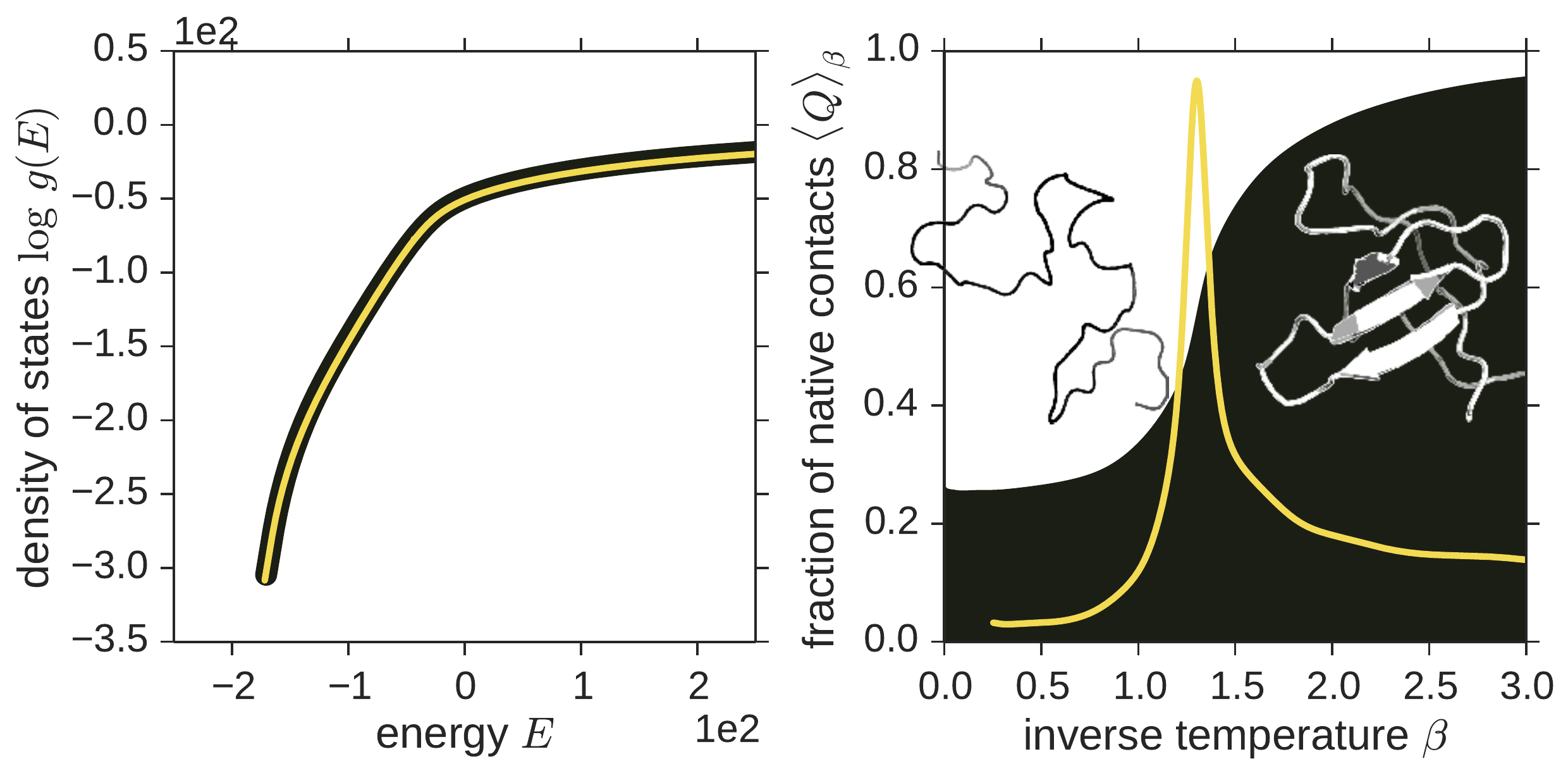}}}
\caption{(Color online) Ensemble annealing of the G{\=o} model. Left panel: Density of states obtained with ensemble annealing (yellow [gray] line) and a parallel tempering simulation (thick black line). Right panel: Fraction of native contacts $\langle Q \rangle_{\beta}$ as a function of the inverse temperature $\beta$ (black area). The heat capacity is indicated by the yellow [gray] line and has been scaled to match the ordinate range. It peaks at the folding temperature $\beta\approx 1.3$. The ribbon diagrams show configurations in the unfolded state (black ribbon on the left) and in the folded state (white ribbon on the right).}\label{fig:go}
\end{figure}
\subsection{Protein model}
Ensemble annealing readily applies to continuous systems such as G{\=o} models that have been used extensively to study protein folding (see e.g. \cite{Whitford09}). In our version of the G{\=o} model, the dihedral angles are the only conformational degrees of freedom; bond lengths and angles are fixed to ideal values. The energy function is comprised of a generic non-bonded energy potential and the G{\=o} term. The non-bonded energy penalizes atom clashes using the same quartic repulsion term as in Ref. \cite{Habeck05}. The G{\=o} term enforces the native structure by imposing a Lennard-Jones potential on the C$\alpha$ distances between residues in contact in the native state. The inverse temperature $\beta$ serves as the control parameter in ensemble annealing runs. As in Ref. \cite{Habeck05}, we used hybrid Monte Carlo \cite{Duane87} for equilibration. We seeded the simulation with 100 random structures and annealed an ensemble of $N=30$ structures; the relative entropy was set to $D=10^{-2}$. For reference, we also ran a parallel tempering simulation of the G{\=o} model using 37 temperatures. The DOS obtained with ensemble annealing was used to optimize the temperatures to produce an exchange rate of 48\%{} on average. 10000 replica transitions were simulated and an estimate of the DOS was obtained by running histogram reweighting.  

We studied the G{\=o} model derived for a small protein domain, the 59 amino-acid Fyn-SH3 domain (PDB code 1SHF). Figure \ref{fig:go}(a) shows the density of states obtained with ensemble annealing and compares it to the reference computed with an exhaustive parallel tempering simulation. The agreement is very high over the entire energy range. In Figure \ref{fig:go}(b) we study the characteristics of the G{\=o} model as revealed by ensemble annealing. Shown is the average number of native contacts $Q$ as a function of the inverse temperature. The folding transition is marked by a sudden increase in the number of native contacts. The heat capacity peaks at $\beta=1.3$ indicating a folding temperature of roughly $0.77$. 
\begin{figure}
\subfigure[\hspace*{0.2cm}Energy histograms]{
\centerline{\resizebox{0.95\columnwidth}{!}{\includegraphics{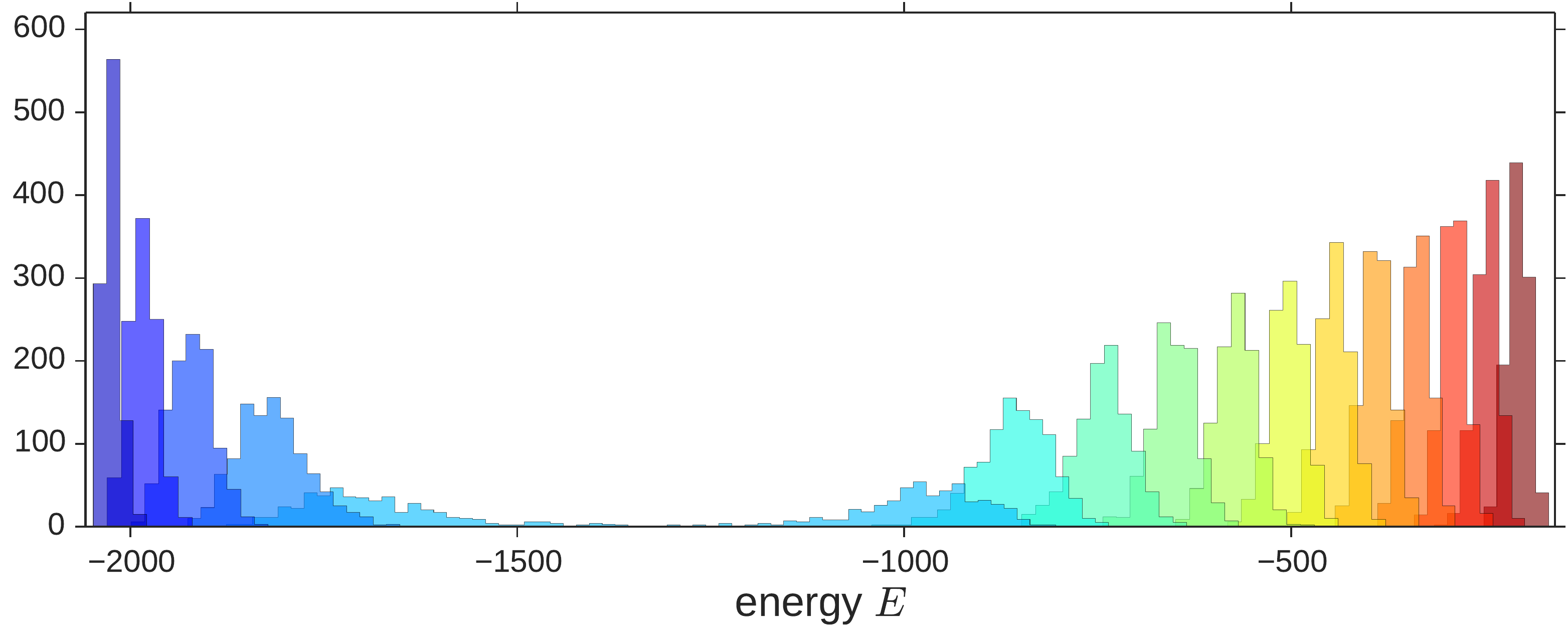}}}
}
\subfigure[\hspace*{0.2cm}Temperature schedule and PT swap rates]{
\centerline{\resizebox{0.995\columnwidth}{!}{\includegraphics{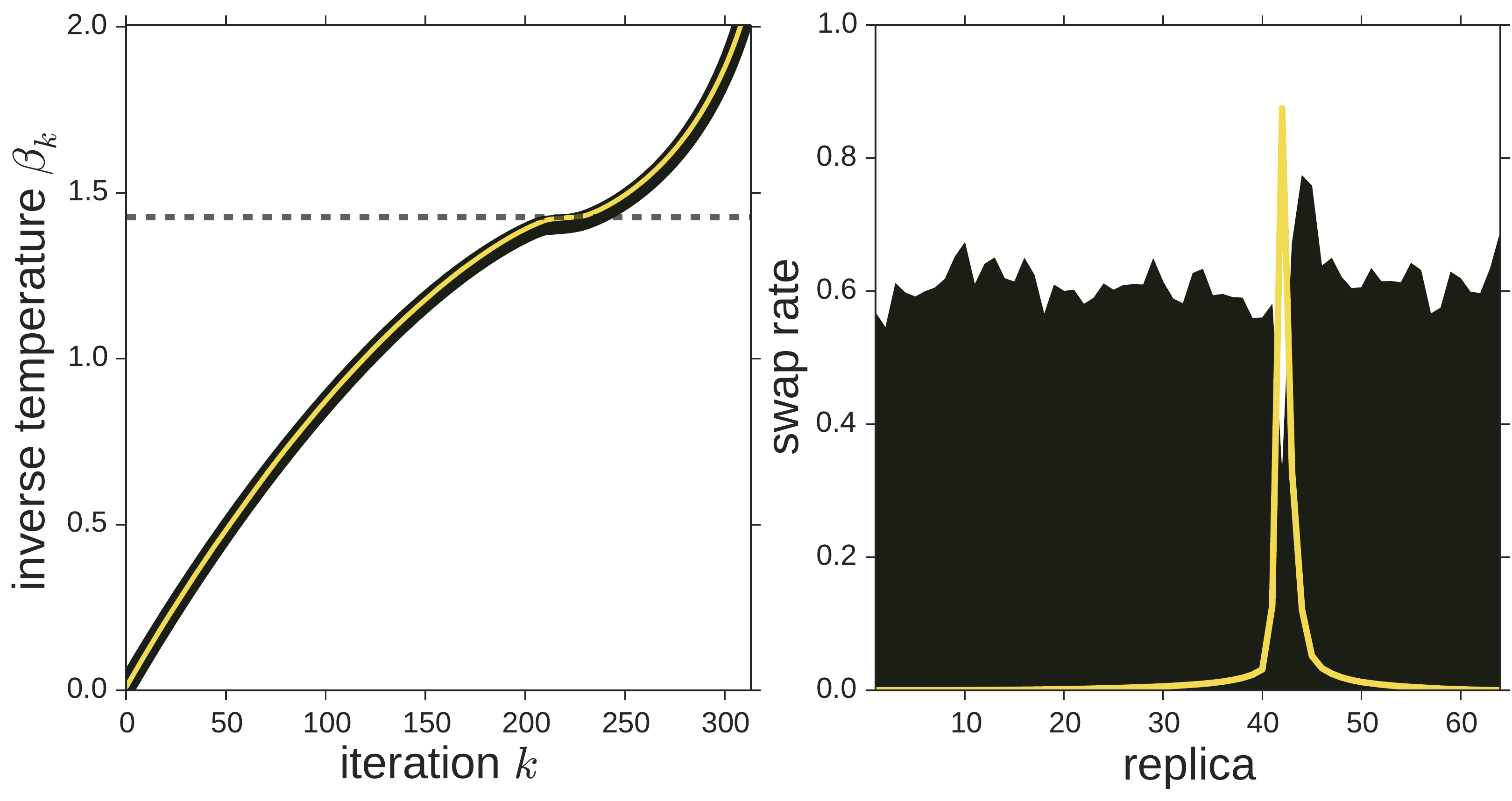}}}
}
\caption{(Color online) Ensemble annealing of ten state Potts model. (a) Energy histograms at the temperatures found by ensemble annealing. Only every twentieth histogram is shown for clarity. The temperature is indicated by the color. (b) Left: Temperature schedule. The thin yellow [gray] line indicates the schedule found by ensemble annealing. The thick black line corresponds to the protocol based on minimum entropy production \cite{Salamon88}. The dashed gray line indicates the critical temperature. Right: PT swap rates obtained with the temperature schedule. The heat capacity is shown as yellow [gray] curve.}\label{fig:2}
\end{figure}
\section{Schedules and paths} 
We will now have a closer look at the schedules constructed by ensemble annealing and compare them to other schedules that have been proposed in the literature. Moreover, we discuss the possibility to use ensemble annealing as a numerical method to construct near-optimal thermodynamic paths.  
\subsection{Temperature schedule}
Figure \ref{fig:2} shows the energy histograms and temperature schedule found by ensemble annealing for the ten state Potts model. By way of construction of the schedule, the energy histograms of successive ensembles have a constant overlap (Fig. \ref{fig:2}(a)). The temperature schedule is non-trivial and deviates from the linear, geometric, and logarithmic schedules that have been proposed in the literature \cite{Kirkpatrick83,Gelman98}. Initially, the inverse temperatures grow sublinearly. In this phase, the schedule constructed by ensemble annealing is reminiscent of the logarithmic schedule proposed by Geman and Geman \cite{Geman84}, i.e. $\beta_k = \beta_0 \ln (1 + k)$. As ensemble annealing approaches the critical temperature, the cooling rate it slowed down automatically such that the system is not quenched and avoids being trapped in a metastable state. Beyond the critical point, the temperatures show a super-exponential increase (Fig. \ref{fig:2}(b)). 

Salamon and co-workers have proposed an adaptive version of simulated annealing more than two decades ago \cite{Salamon88,Ruppeiner91}. Their algorithm finds the temperature schedule by minimizing the entropy production whereupon the temperature changes inversely proportional to the square root of the heat capacity. This rule follows directly from the constant relative entropy criterion. For small changes in inverse temperature, we have 
\begin{equation}\label{eqn:dKL}
\KL{\beta + \delta\beta}{\beta} \approx \frac{1}{2} (\delta\beta)^2 \partial^2_\beta \ln c(\beta) 
\end{equation}
where 
\[
\partial^2_\beta \ln c(\beta) = \langle (E - \langle E \rangle)^2 \rangle_\beta = C(\beta)/\beta^2 
\]
is proportional to the heat capacity $C(\beta)$. If the desired relative entropy $D$ is small, the increment in inverse temperature is
\[
 \delta \beta = \beta \sqrt{2D/C(\beta)}.
\]
Integration over the inverse temperature increments 
\begin{equation}
\int \delta\beta \approx \sum_{l=1}^k \beta_l \sqrt{2D/C_l} 
\end{equation}
generates a schedule that is very close to the one found by ensemble annealing at finite $D$ (Fig. \ref{fig:2}(b)). Comparison with the schedule derived by Salamon \textit{et al.} shows that $\sqrt{D}$ is proportional to the thermodynamic speed of the annealing process. In the context of Bayesian computation, similar, but independent arguments have been put forward by Skilling \cite{Skilling06} who uses $D$ to control the rate of compression as the system moves from the prior to the posterior probability. 

From a practical point of view, an ensemble annealing run can be used to seed a parallel tempering simulation that has a well-balanced schedule and equilibrated initial states. The right panel in Fig. \ref{fig:2}(b) illustrates that the exchange rates are indeed uniform for a PT simulation when using every fifth temperature of the ensemble annealing schedule. A drop in the swap rate is only observed close to the critical temperature where the heat capacity peaks.

\subsection{Minimal dissipation paths}
Let us now see if the results of the previous section generalize to multiple temperatures. Although  ensemble annealing can be applied to any family of bridging distributions [Eq. (\ref{eqn:family})], let us focus on parametric families of the form
\[
p(x|\lambda) = \frac{1}{c(\lambda)}\, q[E(x); \lambda]\, \pi(x) 
\]
where $\lambda$ denotes the vector of all control parameters. The second order expansion of the relative entropy is \cite{Crooks07b}:
\begin{equation}\label{eqn:KLexpansion}
\KL{\lambda'}{\lambda} \approx \frac{1}{2} (\lambda'-\lambda)^T I(\lambda)\, (\lambda'-\lambda)
\end{equation}
where the zero and first order term vanish because $\KL{\lambda}{\lambda} = 0$ and $\lambda'=\lambda$ is the global minimum of $\KL{\lambda'}{\lambda}$ viewed as a function of $\lambda'$. Because the Fisher information matrix
\begin{equation}
I(\lambda) = \int [\nabla_\lambda \ln p(x|\lambda)][\nabla_\lambda \ln p(x|\lambda)]^T\, p(x|\lambda)\, \dd{x}
\end{equation}
is positive definite, it defines a metric on the space of distributions parameterized by $\lambda$. Equation (\ref{eqn:dKL}) is a special case of the general relation (\ref{eqn:KLexpansion}) for the Boltzmann ensemble with a single temperature, where the Fisher information is simply $\partial^2_\beta \ln c(\beta)$.

In statistics, the Fisher metric has been studied since the beginnings of information geometry. The Fisher information can also be used to define a thermodynamic length and action (see \cite{Crooks07b} and references therein). Quasistatic processes that switch between two thermodynamic states follow minimal dissipation paths in $\lambda$ space. These can be computed by minimizing the thermodynamic length (see, for example, \cite{Diosi96,Crooks07b,Sivak12}). Therefore, the optimal path is a geodesic on the Riemanian manifold equipped with the Fisher information metric. Very similar results have been presented by Gelman and Meng in their work on bridge and path sampling \cite{Gelman98}. 

By taking constant but finite steps in relative entropy followed by an equilibration, ensemble annealing approximates a quasistatic process. After $K$ successful equilibrations, the relative entropy accumulated during ensemble annealing is
\begin{equation}\label{eqn:length}
K D = \sum_{k=0}^{K-1} \KL{\lambda_{k+1}}{\lambda_{k}} \approx \frac{1}{2} \int \dot{\lambda}^T I(\lambda) \dot{\lambda} \, \dd{t}
\end{equation}
and approximates the thermodynamic action due to Eq. (\ref{eqn:KLexpansion}). If we aim to optimize the use of computing resources, we have to minimize $K$, the number of bridging distributions. For a single control parameter this is straightforward: we have to follow the geodesic towards the destination ensemble. In the canonical ensemble, for example, if the destination temperature is lower than the initial temperature (annealing), we have to increase $\beta$ such that $\beta_{k+1} > \beta_k$ also for all intermediate temperatures. For ensembles with multiple control parameters the situation is more complicated because minimizing the accumulated relative entropy [Eq. (\ref{eqn:length})] requires the computation of a discrete geodesic. However the DOS is generally unknown, and we can compute $I(\lambda)$ only in the vicinity of the current state. It is not possible to evaluate reliably the length of an entire path connecting the initial and the destination ensemble. We can only search locally without any guarantee that the generated path is close to the geodesic. 

\section{Non-Boltzmann ensembles}
\begin{figure}
\centerline{\resizebox{0.9\columnwidth}{!}{\includegraphics{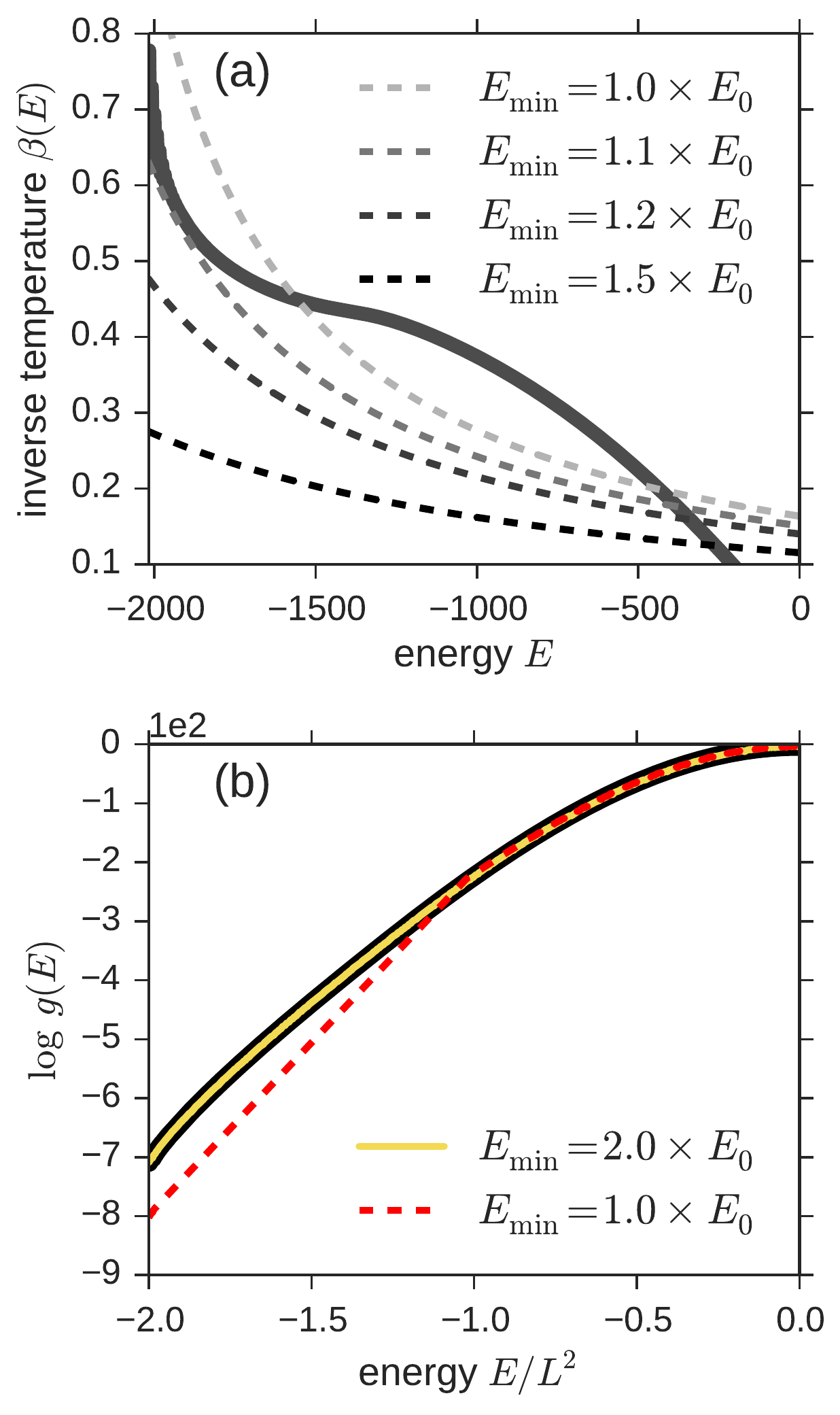}}}
\caption{(color online) Annealing of the $32\times 32$ Ising model in the Tsallis ensemble. (a) Probing different $E_{\min}$ at fixed $\alpha = 1.0025$. The thick black line shows the microcanonical temperature $\beta(E)$. The dashed lines show the right hand side of Eq. (\ref{eqn:tsallis_modes}) for different $E_{\min}$. The curve produced with $E_{\min}$ set to the ground state energy $E_0$ intersects $\beta(E)$ twice corresponding to two peaks in the energy distribution $g(E) q(E; \alpha)$. (b) Comparison of ensemble annealing of the Tsallis ensemble with $E_{\min} = E_0 = -2L^2$ (dashed red line) and $E_{\min} = 2E_0 = -4L^2$ (yellow solid line). The true DOS is shown as thick black line.}\label{fig:tsallis}
\end{figure}
\subsection{Tsallis ensemble}
A major advantage of using the Boltzmann distribution (\ref{eqn:canonical}) as bridging family is that many powerful methods to simulate the canonical ensemble exist. We can use these algorithms in the equilibration step. But it can be beneficial to consider also other ensembles, because they might bridge more efficiently between the initial and final ensemble. The Tsallis ensemble has been used previously in combination with parallel tempering \cite{Hansmann97,Habeck05}. The motivation for this choice is that due to the heavier tails of the Tsallis ensemble replicas have a larger overlap and can exchange states even if they show large energy differences. As a consequence, the number of intermediate replicas should be smaller than with the Boltzmann ensemble.  

This is indeed confirmed by an analysis of the $32\times 32$ Ising model. Test calculations based on the correct DOS show that the canonical ensemble requires 273 $\beta_k$ to reach the destination ensemble ($\beta_\mathrm{final} = 1$) at a relative entropy of $D=10^{-2}$, whereas the Tsallis ensemble needs only 85 $\alpha_k$ values to bridge between $\alpha_\mathrm{start}=1.06$ (corresponding to a very high canonical temperature) and $\alpha_\mathrm{final} = 1.0$. However, in practice this apparent advantage does not hold up. The reason is that the Tsallis ensemble typically yields multimodal energy distributions at intermediate $\alpha_k$. To see this let us first consider the more general case where a parametric bridging family $q(E;\lambda)$ is used. According to Eq. (\ref{eqn:ensemble_energy}) the energy distribution at $\lambda$ is proportional to $g(E) q(E; \lambda)$ and peaks at $\hat{E}$ solving:
\[
0 = s'(\hat{E}) + \frac{q'(\hat{E};\lambda)}{q(\hat{E};\lambda)} = \beta(\hat{E}) + \frac{q'(\hat{E};\lambda)}{q(\hat{E};\lambda)}
\]
where $s(E) = \ln g(E)$ and $\beta(E) = s'(E)$ are the microcanonical entropy and inverse temperature. In case of the canonical ensemble, this equation is simply $\beta(\hat{E}) = \beta$, that is the energy distribution peaks at the energy $\hat{E}$ whose microcanonical temperature matches the canonical temperature. In case of the Tsallis ensemble (\ref{eqn:tsallis}), we have:
\begin{equation}\label{eqn:tsallis_modes}
\beta(\hat{E}) = \frac{\beta}{1 + \beta (\alpha-1) (\hat{E} - E_{\min})}
\end{equation}
This equation can have multiple solutions depending on $\alpha$ and $E_{\min}$, which is why it is difficult to get annealing of the Tsallis ensemble running in a stable fashion. Not only the control parameter $\alpha$, but also the minimum energy $E_{\min}$ plays a critical role (see Fig. \ref{fig:tsallis}). If $E_{\min}$ is exactly set to the energy of the ground state $E_0$, the energy distribution of the Ising model becomes bimodal with a sharp peak around the ground state energy and a second peak corresponding to high temperatures. That is, in order to generate samples from this ensemble we have to simulate two phases simultaneously. As a consequence the DOS estimate produced by ensemble annealing shows systematic errors (Fig. \ref{fig:tsallis}(b)), despite producing an efficient schedule with 103 bridging distributions. If we lower $E_{\min}$, the phase separation is less dramatic and consequently the DOS estimate is as accurate as with the Boltzmann ensemble. But we also lose the efficiency of the Tsallis ensemble in bridging large energy differences, which is reflected in the larger number of $\alpha_k$: 230 $\alpha_k$ for $E_{\min} = 2E_0$ which is similar to the 270 temperatures produced by Boltzmann annealing. This shows that $E_{\min}$ is an additional algorithmic parameter which is delicate to choose. 

\subsection{Microcanonical ensemble and nested sampling}
Nested sampling has been invented by Skilling \cite{Skilling06} to solve Bayesian inference problems. Bayesian inference demands that we draw from a posterior distribution $p(x)$ and compute its normalization constant, which are essentially the tasks that ensemble annealing addresses. In Bayesian inference $\pi(x)$ is the prior, $L(x) = e^{-E(x)}$ the likelihood function; the destination ensemble that we aim to characterize is the posterior distribution over some inference parameter(s) $x$. 

Nested sampling is based on the microcanonical ensemble $q(E; \epsilon) = \Theta(\epsilon - E)$ [Eq. (\ref{eqn:nested})]; the control parameter $\epsilon$ is the maximum energy that the system is allowed to reach \cite{Habeck15}. Therefore nested sampling can be viewed as a special case of ensemble annealing based on a zero-temperature Fermi or the microcanonical ensemble. The relative entropy between two ensembles [Eq. (\ref{eqn:KL})] with energy levels $\epsilon' \le \epsilon$ simplifies to:
\begin{equation}\label{eqn:KLnested}
\KL{\epsilon'}{\epsilon} = \ln[ c(\epsilon) / c(\epsilon')]
\end{equation}
where the normalization constant
\begin{equation}
c(\epsilon) = \int_{E_{\min}}^{\epsilon} g(E)\, \dd{E}
\end{equation}
is the cumulative DOS or configuration space volume. {From a Bayesian point of view,  $c(\epsilon)$ is the prior mass enclosed by the likelihood contour $L = e^{-\epsilon}$.} The control parameter is reduced from infinity to the energy of the ground state $E_{\min}$. 

There are several differences between nested sampling and annealing of the ensemble (\ref{eqn:nested}) using $\epsilon$ as control parameter. These differences result from the fact that all of the features that ensemble annealing aims to implement explicitly are built-in to nested sampling. In fact, nested sampling's design principles served as a guide to develop the ensemble annealing algorithm. 

Ensemble annealing uses histogram methods to estimate the DOS, whereas nested sampling utilizes {\em order statistics} due to the special form the of truncated ensemble (\ref{eqn:nested}). As a consequence of the truncation, $c(E)$ will be uniformly distributed over $p_k(E)$  (defined for $E< \epsilon_k$), which is clear from Eq. (\ref{eqn:ensemble_energy}) \footnote{In statistics, the transformation from $E$ to $c(E)$ where $c(E)$ is the cumulative distribution function of $p(E)$ is called {\em probability transform}. It is a basic mathematical fact that if $E$ follows $p(E)$, $c(E)$ will be uniformly distributed over $[0,1]$.}. Therefore the configuration space volume associated with the maximum energy state follows the distribution $c(E_{\max}) / c(\epsilon) \sim   N t^{N-1}$ where $0 \le t \le 1$ and $E_{\max} < \epsilon$ is the maximum energy among all $N$ walkers. Based on this result from order statistics, nested sampling estimates $c(E)$ at well-dispersed energy cutoffs $\epsilon_k$. 

Another elegant feature of nested sampling is that if $D=1/N$, the next ensemble achieving a compression of $D$ is the one in which the energy is bounded by $E_{\max}$. This results from the fact that $\langle D(E_{\max}||\epsilon) \rangle = - \langle \log t \rangle = 1/N$ where the average is over the Beta distribution $N t^{N-1}$. Therefore the search for the next control parameter will simply yield $\epsilon_{k+1} = E_{\max}$, and we only have to resample the state with the highest energy. 

\begin{figure}
\centerline{\resizebox{0.9\columnwidth}{!}{\includegraphics{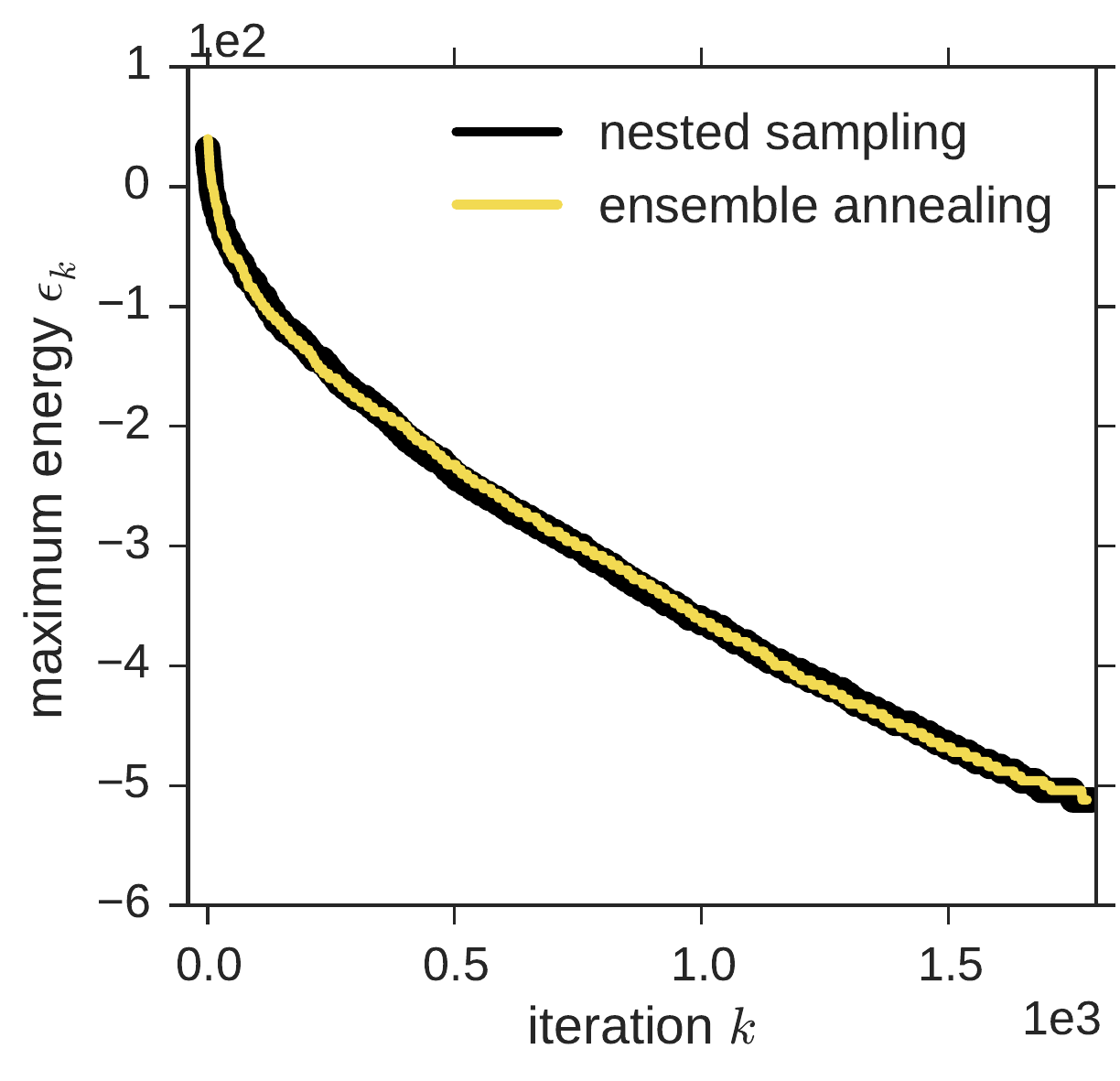}}}
\caption{(Color online) Energy contours $\epsilon_k$ found by nested sampling (thick black line) and ensemble annealing (yellow [gray] line) for the $16\times 16$ Ising model.}\label{fig:ns}
\end{figure}
Although nested sampling is much more efficient at cooling the truncated ensemble (\ref{eqn:nested}), it is also possible to run the ensemble annealing algorithm. Both methods produce comparable sequences of energy levels $\epsilon_k$ for the $16\times 16$ Ising model with $N=10$ and $D=1/N=0.1$ (see Fig. \ref{fig:ns}). Also the estimated DOS is of similar accuracy. For this example, nested sampling runs at a speed that is three orders of magnitude faster than ensemble annealing. This is due to the fact that DOS estimation and annealing (i.e. the choice of the next energy limit) are instantaneous in nested sampling, because they are built-in to the method. Ensemble annealing, on the contrary, needs to run the histogram iterations for every energy contour. The histogram iterations converge only very slowly. Each iteration is dominated by DOS estimation because equilibration of the Ising model is very fast. For other systems such as proteins it will be the equilibration step rather than DOS estimation that consumes most of the computation time. In this situation, the discrepancy between nested sampling and ensemble annealing will not be as dramatic as for the Ising model. 

For the $d$-dimensional harmonic oscillator we have $g(E) = \tfrac{2}{d} E^{d/2-1}$ and $c(E) = E^{d/2}$. As in the canonical ensemble, the relative entropy depends on the ratio of two successive control parameters: $\KL{\epsilon}{\epsilon'} = d/2 \ln(\epsilon'/\epsilon)$. Therefore nested sampling and ensemble annealing progress geometrically according to $\epsilon_k = \epsilon_0 \, \rho^{k}$ where $\rho = e^{-2D/d}$. Let us compare this to the thermal approach using the inverse temperature as a control parameter. The compression rate of the canonical distribution is given by $\rho - \ln\rho = 2D/d+1$ [Eq. (\ref{eqn:canonicalrate})]. Therefore $\rho_\beta - \ln\rho_\beta = -\ln\rho_\epsilon + 1$ where $\rho_\beta$ and $\rho_\epsilon$ are the compression rates of thermal and microcanonical annealing. Rewritten we have $\ln(\rho_\beta/\rho_\epsilon) = \rho_\beta - 1 \le 0$, and therefore $\rho_\beta \le \rho_\epsilon \le 1$. This means that annealing the canonical ensemble compresses faster than annealing the microcanonical ensemble. We observe this for the application to the Ising model (Fig. \ref{fig:ns}). Canonical annealing with $N=10$ walkers and a relative entropy of $D=0.1$ requires only 42 iterations to reach the destination ensemble. Microcanonical annealing and nested sampling, on the contrary, need approximately 1800 iterations until convergence. The reason for this is that states accumulate at the maximum energy $\epsilon$, and therefore nested sampling and microcanonical annealing will produce many intermediate ensembles in order to bridge between the initial and the destination ensemble. 

\section{Conclusion}
Ensemble annealing is a Monte Carlo algorithm that steps through a sequence of ensembles and generates conformational samples. Along with the samples, it also estimates the density of states using histogram methods. The ensembles are placed in an adaptive manner so as to maintain a constant, pre-chosen relative entropy between successive ensembles. Ensemble annealing can be applied to a variety of bridging distributions, foremost the canonical ensemble but also to non-Boltzmann families such as the Tsallis or the microcanonical ensemble. There is a close connection to the nested sampling algorithm. In fact, ensemble annealing aims to implement the features that are built-in to nested sampling: control of the compression or thermodynamic speed, as well as reliable estimation of the compression based on the DOS or the configuration space volume. Nested sampling is intimately tied to the truncated ensemble (\ref{eqn:nested}), whereas ensemble annealing is more general in the choice of the ensemble itself, which can help to speed up the simulation and allows the use of samplers that work efficiently with a particular ensemble (such as, for example, hybrid Monte Carlo in the canonical ensemble). 

Ensemble annealing is also related to previous work by Salamon and co-workers \cite{Salamon88,Ruppeiner91} on simulated annealing. Our approach is more general and gives richer results because it not only finds the system's ground state but reconstructs the entire DOS. That way ensemble annealing can be used to both simulate thermodynamic systems and solve difficult optimization problems. By means of the DOS, all visited states contribute to the computation of ensemble averages making our approach more robust. Moreover, it is possible to work with multiple control parameters, which will be studied in future extensions of ensemble annealing.

\begin{acknowledgments}
This work was supported by Deutsche Forschungsgemeinschaft (DFG) grants No. HA 5918/1-1 and No. SFB860 TP B9.
\end{acknowledgments}
\vspace*{-0.4cm}
%

\end{document}